\documentclass{kluwer}    
\usepackage[dvips]{graphicx}

\newdisplay{guess}{Conjecture}

\begin{document}
\begin{article}
\begin{opening}
\title{Gamma-ray burst interaction with dense interstellar
medium}
 \author{Maxim V. \surname{Barkov}}
 \author{Gennady S.  \surname{Bisnovatyi-Kogan}}
 \runningauthor{Maxim V. \surname{Barkov} and
  Gennady S.  \surname{Bisnovatyi-Kogan}}
 \runningtitle{Gamma-ray burst interaction with dense interstellar
medium}
 \institute{Space Research Institute (IKI) Russian Academy of Science}
 \date{\today}

\begin{abstract}
Interaction of cosmological gamma ray burst radiation with the
dense interstellar medium of host galaxy is considered. Gas
dynamical motion of interstellar medium driven by gamma ray burst
is investigated in 2D approximation for different initial density
distributions of host galaxy matter and different total energy of
gamma ray burst. The maximum velocity of motion of interstellar
medium is $1.8\cdot10^4$ km/s. Light curves of gamma ray burst
afterglow are calculated for set of non homogeneous density,
distribution gamma ray burst total energy, and different viewing
angles. Spectra of gamma ray burst afterglow are modeled taking
into account conversion of hard photons (soft X-ray, hard UV) to
soft UV and optics photons.
\end{abstract}
\keywords{gamma ray burst, optical afterglow, }

\end{opening}

\section{Introduction.}

Although gamma-ray bursts (GRBs) were discovered more then thirty
years ago \cite{Kleb73}, their origin is still unclear. The most
extensive data on the detection of GRBs have been obtained by the
Compton Gamma Ray Observatory BATSE experiment \cite{Briggs95,
Fishman95,Meegan92}. Analysis of the GRBs detected showed that
their apparent distribution on the sky was isotropic, but that
there was a significant departure of their (logN -- logS) curve
from the $N\sim S^{-3/2}$, low corresponding to a spatially
uniform source distribution \cite{Briggs95,Kouv94}.

Observation of optical afterglows  of GRB, following after
identification of GRB with a transient X-ray source by Beppo-SAX,
and discovery of large (up to $z=4.5$) redshifts in the spectra of
optical transients had confirmed the cosmological origin of long
GRB.

The cosmological model suggests that the GRB source are located in
distant galaxies (within $\sim 10^3$ Mpc). In the framework of
this model, the observed fluxes $\sim 10^{-4}$ erg s$^{-1}$
require the release of enormous amounts of energy ($\sim 10^{51} -
10^{53}$ erg) within a fairly short time interval (of the order of
several tens of seconds). Such a powerful energy release should
have a strong effect on a large volume of matter around the parent
galaxy and should give rise to the formation of GRB counterpart at
other wavelengths. Without specifying the mechanism for GRB
formation, we assume only the existence of a strong flux gamma
radiation and consider the interaction of this radiation with the
interstellar medium on large spatial and temporal scales.

We investigate the response of an interstellar medium of standard
chemical composition to the passage of a short-term powerful pulse
of gamma radiation (that is, the dynamical behaviour and radiative
cooling after heating by the gamma rays).

The spherical symmetric model was investigated by Bisnovatyi-Kogan
and Timokhin (1997), 2D model, permitting to study different
matter distribution and reaction to anizotropic GRB had been
studied by Barkov and Bisnovatyi-Kogan (2004a, 2004b) using
numerical simulations by PPM method. In what follows we represent
the results from Barkov and Bisnovatyi-Kogan (2004a, 2004b).

\section{The main equations}

We solve the system of hydrodynamic equations, describing the
motion of matter, together with thermal processes in axially
symmetric case

\begin{equation}\label{uog1}
   \frac{\partial \rho}{\partial t}+\nabla(\rho \vec{v})=0,
\end{equation}

\begin{equation}\label{uog2_r}
   \frac{\partial (\rho v_{r})}{\partial t}
   + \frac{\partial( \rho v_{r}^2+P)}{\partial r}
   + \frac{1}{ r}\frac{\partial( \rho v_{r} v_{\theta})}{\partial
\theta}
   + \frac{2 \rho v_{r}^2 - \rho v_{\theta}^2 +
     \rho v_{r} v_{\theta} \: ctg \:\theta}{ r}
       =
   \rho {F_{\gamma}},
\end{equation}

\begin{equation}\label{uog_e}
    \frac{\partial (\rho v_{\theta})}{\partial t}
   + \frac{\partial (\rho v_{r} v_{\theta})}{\partial r}
   +  \frac{1}{ r}\frac{\partial ( \rho v_{\theta}^2+P)}{\partial
\theta}
   + \frac{3 \rho v_{r} v_{\theta} +
     \rho v_{\theta}^2\: ctg \:\theta}{ r}
    = 0,
\end{equation}

\begin{equation}\label{uog3}
   \frac{\partial }{\partial t}\left( \frac{\rho v^2}{2} + \rho\varepsilon\right)
   +\nabla\left\{ \rho\vec{v}\left( \frac{v^2}{2}
   +\varepsilon+ \frac{P}{\rho} \right) \right\}=
   \rho H_{\gamma}-\rho C_{\gamma}.
\end{equation}

Here $\rho, P, \varepsilon, v_r,$ and $v_{\theta}$ are density,
pressure, internal specific energy and two velocity components,
respectively. The gamma ray pulse is considered as instant one
having total energy $\Gamma$ and luminosity
\begin{equation}\label{uog5}
    L=\Gamma \delta \left(t-\frac{r}{c}\right).
\end{equation}

We are interested in the behaviour of the gas heated by GRB, which
is taken as fully ionized one.

Consider flat spectrum of GRB

\begin{equation}\label{uog5a}
    \frac{dL}{dE} = \frac{L}{E_{max}} e^{-E/E_{max}}
\end{equation}

The main GRB photons have energies larger then ionization energies
of most electrons, so we consider energy exchange of GRB with the
gas due to Compton and inverse Compton processes only. The
function $H_{\gamma}$ in (\ref{uog3}) is written us
\begin{equation}\label{uog6}
    H_{\gamma}=\frac{L}{4\pi r^2}\frac{\mu_e \sigma_T}{m_u}
    \frac{E_{max}f_h(E_{max})-4kT f_c(E_{max})}{m_e c^2},
\end{equation}
where
 $$
    f_c(E_{max})= \int^{\infty}_{0} W(E,E_{max}) q(E) dE,
 $$
\begin{equation}\label{b_i_1}
    f_h(E_{max})= \frac{1}{E_{max}}\int^{\infty}_{0}
    W(E,E_{max}) s(E) E dE.
\end{equation}
We have used GRB spectra with $E_{max}=0.6$ MeV and 2 MeV. The
functions $s(E)$ and $q(E)$, which take into account deviations
from Thomson cross section $\sigma_T$ due to Klein-Nishina
correction ($\sigma_{KN}$), are taken from (Beloborodov and
Illarionov, 1995). They are normalized so that $s(E) = q(E) = 1$
at $E\ll m_e c^2$. In our cases $f_h=0.19; f_c=0.33$ for $E=0.6$
MeV, and $f_h=0.065; f_c=0.16$ for $E=2$ MeV.

The radiation force due to electron scattering is written as
\begin{equation}\label{uog4}
  F_{\gamma}= \frac{1}{c}\frac{L}{4\pi r^2}\frac{\mu_e
  \sigma_T}{m_u} f_f(E_{max}),
\end{equation}
where the function
\begin{equation}\label{b_i_3}
    f_f(E_{max})= \frac{1}{\sigma_{T}}\int^{\infty}_{0}
    W(E,E_{max}) \sigma_{KN}(E) dE
\end{equation}
takes into account KN corrections, $f_f=0.5$ for $E_{max}=0.6$ MeV
and $f_f=0.32$ for $E_{max}=2$ MeV. Cooling of the gas is due to
different radiative processes (ff, fb, bb). For optically thin
plasma,  which is used for description of cooling, the function
$C_{\gamma}$ was calculated in Kirienko (1993), Raymond, Cox and
Smith (1976)
\begin{equation}\label{uog7}
    C_{\gamma}=\frac{\Lambda (T) n^2}{\rho},
\end{equation}
where $ \Lambda (T)$ from  Kirienko (1993) was approximated
analytically with a precision not worse then 5\%
\begin{equation}\label{uog8}
    \Lambda (T) =
    \left\{
        \begin{array}{ll}
        0, & T<10^4 K\\
        10^{-48.8}T^{6.4},  & 10^4<T<10^{4.25}\\
        10^{-16.5}T^{-1.2},  & 10^{4.25}<T<10^{4.5}\\
        10^{-27.48}T^{1.24},  & 10^{4.5}<T<10^{5}\\
        10^{-21.03}T^{-0.05},  & 10^{5}<T<10^{5.4}\\
        10^{-13.6698}T^{-1.413},  & 10^{5.4}<T<10^{5.86}\\
        10^{-22.8378}T^{0.1515},  & 10^{5.86}<T<10^{6.19}\\
        10^{-13.1969}T^{-1.406},  & 10^{6.19}<T<10^{6.83}\\
        10^{-22.2877}T^{-0.075},  & 10^{6.83}<T<10^{7.5}\\
        10^{-26.6}T^{0.5},  & 10^{7.5}<T
        \end{array}
    \right.
\end{equation}
It was shown in Barkov and Bisnovatyi-Kogan (2004a) that the heat
conductivity may be neglected in this problem.

\section{Numerical results}

In the paper of Barkov and Bisnovatyi-Kogan (2004b) more then 10
variants with different density distribution and GRB beaming have
been calculated. Here we represent the main results of these
calculations.

\begin{figure}
 \centerline{\includegraphics[width=3.5in]{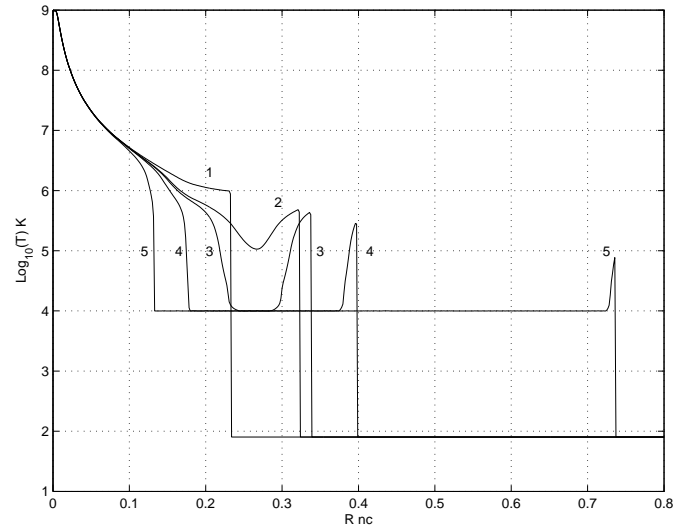}}
 \caption[]{ The evolution of temperature distribution in
 the cloud with time is represented for the GRB
burst in the center of a spherically symmetric uniform cloud with
a radius $R=1.5$ pc, concentration $n_H=10^5$ cm$^{-1}$,
$\Gamma=10^{52}$ erg, $E_{max}\geq 1/6 $ MeV;  curves marked by
number correspond to the following time moments after GRB: 1) 0.76
year, 2) 1.054 year, 3) 1.103 year,
 4) 1.30 year, 5) 2.40 year.}
 \label{Figure_40}
\end{figure}

In the fig. 1 from Barkov and Bisnovatyi-Kogan (2004b) the
evolution of temperature distribution in the cloud with time is
represented for the GRB exploding in the center of a spherically
symmetric uniform cloud with a radius $R=1.5$ pc, concentration
$n_H=10^5$ cm$^{-1}$, $\Gamma=10^{52}$ erg, $E_{max}\geq 1/6 $
MeV. It was shown by Barkov and Bisnovatyi-Kogan (2004b), that for
$E_{max}\geq 1/6 $ MeV, the heating at big distance from GRB
($r\geq 0.05$ pc) depends only on the GRB energy $\Gamma$ and does
not depend on $E_{max}$. The temperature inversion is developed in
the middle radiuses ($r=0.15\div 0.7$ pc), where after heating
$T\sim 10^6$ K, and the cooling is the most effective. The cooling
front is propagating outward with superlight speed, and inward
with sublight speed (phase velocities). The light curve for
optical and ultraviolet luminosity, observed by the distant
observer is represented in fig. 2 from Barkov and Bisnovatyi-Kogan
(2004b).
\begin{figure}[tbp]
 \includegraphics[width=3.5in]{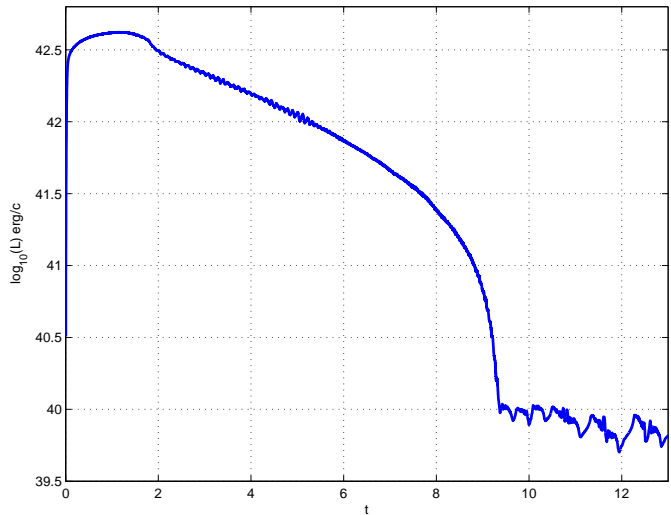}
 \caption{  The light curve for optical and ultraviolet
luminosity for same parameters, as in fig. 1, observed by the
distant observer. Time is in years.}
 \label{Figure_29}
\end{figure}

GRB heating of the cloud is most intensive in central parts and
leads to formation of the shock wave propagating outwards. The
speed of the shock is about $2\times10^8$ cm/s for the uniform
spherical cloud, what is much less then the light speed.
Therefore, the cloud is heated mainly by the light signal from
GRB, and effects connected with the formation of central shock do
not influence the integral light curve, except the radiation in
the hard X-ray band produced in close vicinity of GRB explosion.

In the case, when GRB explosion takes place between two dence
clouds, or in the cavity, produced by strong anizotropic stellar
wind, the hydrodynamic effects may be much stronger then in the
case of the uniform cloud, and the speed of the shock increase
upto $\sim 2\times10^9$ cm/s.

\begin{figure}[tbp]
 \includegraphics[width=3.5in]{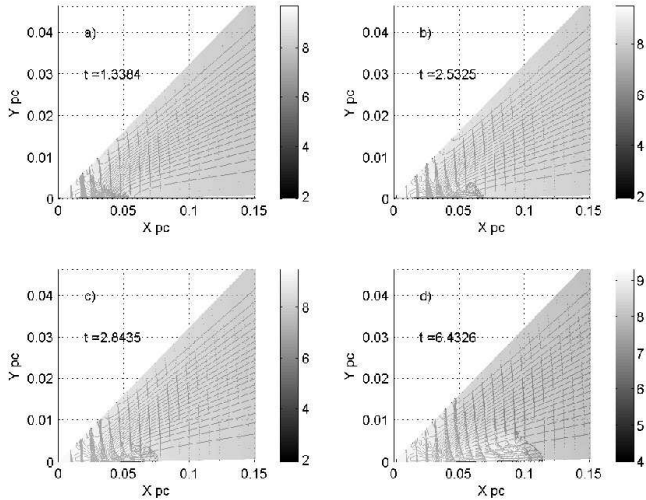}
 \caption{ the evolution of the velocity
field is represented for the conical density distribution in the
cloud.}
 \label{Figure_44}
\end{figure}

In the fig. 3 from Barkov and Bisnovatyi-Kogan (2004a) the
evolution of the velocity field is represented for the case of the
density distribution in the cloud as
$n=10^5e^{\displaystyle{-2-2\cos(10\theta)}}$cm$^{-3}$, cloud
radius 1.5 pc, for the energy of isotropic GRB
$\Gamma=1.6\times10^{53}$ erg, and $E_{max}\geq 1/6$ MeV. The
calculation have been performed in the region $0\leqslant \theta
\leqslant \pi/10$, with the condition $v_{\theta}=0$ on the
boundaries. The temperature of the heated gas depend only on the
distance from GRB, so pressure gradient is developed inside the
cavity inducing the motion of the matter to the axis of the cone.
Collision of the flow at the cone axis produces a cumulative
effect, leading to matter acceleration along the axis up to
velocity $\sim 2\times10^9$ cm/s Barkov and Bisnovatyi-Kogan
(2004a).

The accelerated matter in this case have a form of the bullet. In
the case of the explosion in the space between two spherical
clouds, the ejected matter should have the form of an expanding
ring.

\begin{figure}[tbp]
 \includegraphics[width=3.5in]{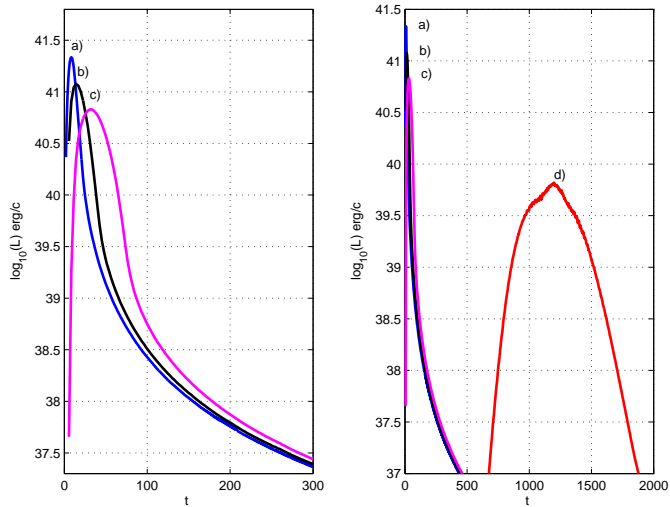}
 \caption{ Light curves for collimated GRB,situated  on 1 pc
 from center of the  molecular cloud, time in years.
 Observer's situated on the line GRB - MC, case a);
 deviation from this line by the angle $\alpha = 0.1$ radian, case b);
 by the angle $\alpha = 0.2$ radian, case c);
 by the angle $\alpha = \pi/2$ radian, case d).}
 \label{Figure_s3a}
\end{figure}

In the case of anizotropic GRB exploded in the uniform gas cloud
the observed light curve is different for distant observer with
different angular distances from the symmetry axes. Such light
curves are represented in fig 4. from Barkov and Bisnovatyi-Kogan
(2004b). Here anizotropic GRB is considered with angular
dependence of luminosity $\Gamma(\theta) =
10^{52}e^{\displaystyle{-\left({\vartheta}/{\vartheta_0}\right)^2}}$,
$\theta=0.1$ rad, the total energy of GRB
$\Gamma_{tot}=2.5\times10^{49}$ erg, density distribution
$\displaystyle{ n=10^{5} e^{\displaystyle{
-\left({r}/{r_0}\right)^2 }} }$ cm$^{-3}$, $r_0=0.2$ pc, when the
explosion take place at the distance 1 pc from the center of the
cloud. The shortest optical burst of few days with largest
luminosity $\sim 10^{41.5}$ erg/s is seen by the observer,
situated at the symmetry axis on the continuation of the line
cloud center --- GRB.

The observer on the line which is perpendicular to the symmetry
axis is observing much longer optical afterglow ($\sim 1000$
days), but with accordingly  lower luminosity. All these
differences  are connected with the kinematic of the light
propagation from the nonuniformly and nonsimultaneously heated gas
cloud.

\section{Discussion}

Optical afterglow, connected with the reradiation  of the GRB by
the dense enough molecular cloud could be observed as optical
transient. Indications that GRB explosions take place in the
region of star formation filled with dense gas clouds
\cite{Sokol,Pacz} make this possibility as very probable.
Observation of plato in the optical afterglow of GRB 030329 during
a month between 64 and 94 days after GRB detection \cite{Poz} may
be connected with such kind of reradiation.

The dense molecular clouds have very low temperature, at which
dust is formed. Estimations made by Barkov and Bisnovatyi-Kogan
(2004b) show, that dust is evaporated by GRB pulse in the cloud
near GRB impulse direction, or in the whole cloud by isotropic GRB
with $R\sim 1$ pc, and does not influence the light curve of
optical, UV and X-ray GRB afterglows.

\end{article}
\end{document}